\documentclass[aps,prb,showpacs,twocolumn,groupedaddress,amsmath]{revtex4}

\usepackage{graphicx}

\newcommand\cpl{Chem.~Phys.~Lett.}
\newcommand\jpca{J.~Phys.~Chem.~A}
\newcommand\jpcm{J.~Phys.~Cond.~Matt.}
\newcommand\lang{Langmuir}

\newcommand\nmat{Nature Materials}

\newcommand\sci{Science}

\begin{document}
\title{Novel Structural Motifs in Clusters of Dipolar Spheres: Knots, Links and Coils}
\author{Mark A.~Miller and David J.~Wales}
\affiliation{University Chemical Laboratory, Lensfield Road, Cambridge CB2 1EW, United Kingdom}
\date{\today}

\begin{abstract}
We present the structures of putative global potential energy minima for clusters bound
by the Stockmayer (Lennard-Jones plus point dipole) potential.  A rich variety of structures
is revealed as the cluster size and dipole strength are varied.  Most remarkable are
groups of closed-loop structures with the topology of knots and links.  Despite the
large number of possibilities, energetically optimal structures exhibit only a few
such topologies.
\end{abstract}

\maketitle

Isotropic van der Waals or depletion forces favor compact packing of spherical
particles, since such arrangements lead to a large number of nearest neighbor
interactions.  For small clusters, icosahedral motifs often prevail,\cite{Wales97a}
giving way to close packing for sufficiently large
or bulk systems.  In contrast, the energetically optimal arrangement of two purely
dipolar particles is with the dipoles aligned head-to-tail, leading to chain
formation.  In systems with both isotropic and dipolar interactions, frustration
arises from the competition between these two effects.
\par
The study of dipolar sphere systems has a long and controversial history, concerned mostly
with the existence and nature of their phase transitions.\cite{Teixeira00a}
Fresh impetus comes from recent experiments, in which nanoparticles of low polydispersity with
a magnetic dipole have been synthesized and studied by electron microscopy.\cite{Butter03a}
Dipolar particles have also been harvested from magnetotactic bacteria.\cite{Philipse02a}
Colloidal systems like these provide great scope for exploring the influence of
particle interactions on structure and dynamics both because highly detailed information
is available from electron or confocal microscopy, and because the form of the interactions
can be finely tuned by the experimental conditions.\cite{Frenkel02a}
\par
Dipolar particles start to aggregate at low volume fractions,\cite{Butter03a} producing
chains and clusters.  In this Letter, we identify the energetically optimal
structures of clusters composed of spherical particles that interact through a permanent
(electric or magnetic) dipole moment in addition to an isotropic soft core and attractive
tail.  We ask how the morphology of the clusters is affected by the number of particles
and the strength of the dipolar interaction relative to the isotropic attraction due to
van der Waals or depletion forces.
\par
We start with the Stockmayer potential,
\begin{multline*}
V=\epsilon\sum_{i<j}^N \Bigg\{4\left[\left(\frac{\sigma}{r_{ij}}\right)^{12}
-\left(\frac{\sigma}{r_{ij}}\right)^6\right]+\\
\frac{\mu^2\sigma^3}{r_{ij}^3}\left[\hat{\boldsymbol{\mu}}_i\cdot\hat{\boldsymbol{\mu}}_j-
\frac{3}{r_{ij}^2}(\hat{\boldsymbol{\mu}}_i\cdot{\bf r}_{ij})
(\hat{\boldsymbol{\mu}}_j\cdot{\bf r}_{ij})\right]\Bigg\},
\end{multline*}
for the total pairwise interaction energy of $N$ dipolar particles, where ${\bf r}_{ij}$ is the position
of particle $j$ relative to particle $i$ and $\hat{\boldsymbol{\mu}}_i$ is a unit vector
along the dipole moment of particle $i$.
The units of energy and length are set by the Lennard-Jones (LJ) $\epsilon$ and $\sigma$ parameters,
while the dimensionless parameter $\mu$ determines the strength of the dipolar contribution
relative to the LJ part.
\par
Considerable progress has been made in the field of global optimization in the decade since
Clarke and Patey first searched for the lowest energy structures of small dipolar structures
using simulated annealing.\cite{Clarke94a}  These authors concentrated particularly on the $N=13$
cluster, denoted here St$_{13}$, finding a sequence of structures with increasing $\mu$ that
has been verified
recently.\cite{Oppenheimer04a}  For larger sizes, ``intestinal'' structures were reported, and
the analysis of these intricate assemblies is one purpose of the present contribution.
We employ the well documented basin-hopping algorithm,\cite{Wales97a} in which a Monte
Carlo simulation is run on a transformed potential energy surface by performing a local
geometry optimization at each step.  We found that geometry optimizations converged more
efficiently if each $\hat{\boldsymbol{\mu}}_i$ is represented by a pair of spherical polar
angles rather than by a Cartesian vector, since the latter contains a redundant third
degree of freedom.
\par
Global optimization runs were performed for $3\le N\le55$ particles and a grid of dipole strengths
covering the range $0\le\mu\le6$, each run initiated from a random configuration.
Starting from the optimal structure of a neighboring size is often detrimental, since the
global minimum can change abruptly with $N$ or $\mu$, potentially trapping the search in
the wrong morphology.  To prevent particles from occasionally drifting away,
the cluster was placed inside
a hard spherical container.  The number and length of runs varied with 
cluster size and dipole strength as required to obtain reliably reproducible global
minima.  Since the dipolar contribution to the energy changes with the square of
$\mu$, the relevant energy scale increases rapidly as one moves away from the $\mu=0$
LJ limit.  Some care must be taken in choosing a suitable temperature $kT$ for the
basin-hopping searches, to ensure that the transformed surface is properly explored, while
the global minimum structure retains significant statistical weight.  We found that
$kT/\epsilon=1+\mu^2/5$ was a useful starting point, though other schemes were tried in
cases of doubt.

\begin{figure}[t]
\includegraphics[width=85mm]{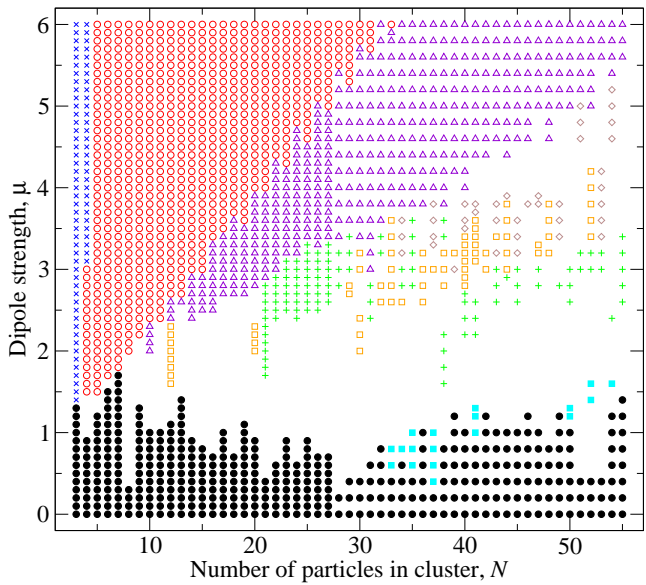}
\caption{Structural map for global potential energy minima of Stockmayer
clusters.  Symbols denote morphologies: relaxed Lennard-Jones (black filled circle), decahedral
(cyan filled square), linear (blue cross), ring (red open circle), two stacked rings (purple
triangle), coil (brown diamond), link (orange open square), knot (green plus).
Structures that do not fall into these categories have been omitted for clarity.
\label{map} }
\end{figure}

Figure \ref{map} summarizes the occurrence of the
most prominent morphologies resulting from global
optimization.  At $\mu=0$, the well known LJ structures are reproduced,\cite{Wales97a}
which, with the notable exception of the face-centred cubic 38-particle
cluster, are constructed from icosahedral motifs.  A small non-zero $\mu$ typically leads
to a slight relaxation of the LJ structure.  The preference of dipoles for head-to-tail
alignment gives rise to frustrated circuits of dipoles within the LJ structure,
lowering the symmetry.\cite{Lavender94a}  For example, the perfect icosahedral symmetry
of the 13-atom LJ cluster is broken by loops of dipoles that point around one
of the three-fold axes.  Even without decorating the particles with the dipole vectors, the
point group defined by the particle positions changes from $I_h$ to $D_3$.

At some point, the dipole contribution to the energy becomes large enough to favor a
different structure.  The threshold at which the change happens varies substantially with
$N$, depending on the nature of the competing structures.  In some cases, such as
$51\le N\le54$, a vacancy in an icosahedral shell simply finds a more favorable site.  In
other cases, an entirely different class of compact structures takes over, such as
for certain sizes in the range $33\le N\le41$, where there is a
switch to a decahedral motif with at least one incomplete shell.  Surprisingly, the
dipoles circulate roughly about an axis perpendicular, rather than parallel, to the
pseudo five-fold axis.  The emergence of decahedra here
is partly explained by the fact that the number of nearest neighbor
pairs in decahedral structures is high, relative to the underlying average, for
odd $N$ close to 38,\cite{Doye95b} making decahedra competitive with icosahedra for
alternate $N$ close to this size.  The exception is $N=39$, where the icosahedral
structure has a stable partial second shell.  This subtle interplay of forces is one of several
examples that give rise to the diverse results mapped out in Fig.~\ref{map}.
\par
\begin{figure}[t]
\includegraphics[width=75mm]{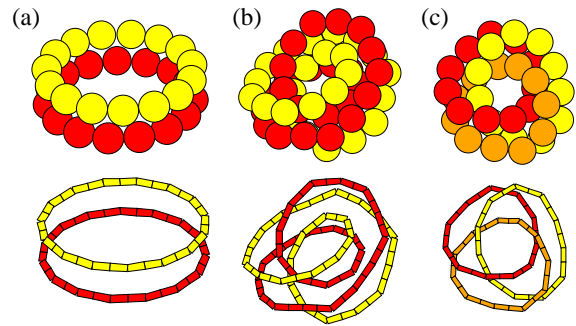}
\caption{Some multi-component global minima. (a) St$_{30}$ at $\mu=3.6$:
the trivial unlink,
(b) St$_{48}$ at $\mu=3.4$: two linked coils, $4^2_1$, and
(c) St$_{33}$ at $\mu=3.2$: the link $6^3_3$.
In each case, spherical particles are shown in the upper panel, and the underlying chain of
dipoles in the lower.  Components within each link are distinguished by color.
\label{links} }
\end{figure}

In the opposite limit of high $\mu$, the global minimum is a ring with the dipoles
tangential to the edge for all $N\ge5$.  Compared with a linear structure, the energy
required to bend the chain of dipoles into a ring is more than compensated by the
additional contact made between the ends.  When $\mu$ is lowered sufficiently for
$N=10$ and $N\ge12$, the
LJ contribution to the energy makes it more favorable to form two rings of
half the size, which then stack on top of each other, creating one new nearest neighbor
per particle.  The threshold of $\mu$ at which this change occurs shows an overall
increase with $N$, since
the difference in bending energy per particle between a ring of $N$ particles and two
rings of $N/2$ particles decreases with $N$.  For even $N$ the two rings are planar,
forming an antiprism, as illustrated in Fig.~\ref{links}(a).  For odd $N$, the two
rings differ in size by one particle, and an out-of-plane distortion is necessary to
accommodate the mismatch.  Since some of the dipole--dipole interactions are frustrated
in this arrangement, the single ring becomes more favorable at a lower $\mu$ for
odd $N$ than for neighboring even $N$.  The odd--even alternation in the boundary
between one and two rings is clearly visible in Fig.~\ref{map}.

The antiprismatic structure in Fig.~\ref{links}(a) shows two clearly defined closed
circuits of dipoles.  While neighbors in the same chain approach somewhat more closely
than members of different rings, making it possible to define chains using a distance
criterion for neighbors, it is much less ambiguous to use an energetic definition.
The anisotropy of the dipolar potential means that the propensity of a particle for
``bonding'' is effectively saturated by two particles with roughly aligned dipoles
positioned at its head and tail.
Starting from a given particle $i$, we define the next member of the chain as the
particle $j$ that has the lowest dipole--dipole interaction energy with $i$
and is located in the half-space to which the dipole of $i$ points,
i.e.~$\hat{\boldsymbol{\mu}}_i\cdot{\bf r}_{ij}>0$.  The previous particle in the chain
is likewise the particle with lowest interaction energy that also satisfies
$\hat{\boldsymbol{\mu}}_i\cdot{\bf r}_{ij}<0$.  This parameter-free definition
reliably and intuitively decomposes a structure into its constituent chains.
In compact clusters, where the
dipole--dipole interactions are not dominant enough for the structure to be
naturally decomposed into chains, an attempt to define chains in this way will
result in a collection of meaningless fragments and can be discarded as
irrelevant.

\begin{figure}[t]
\includegraphics[width=75mm]{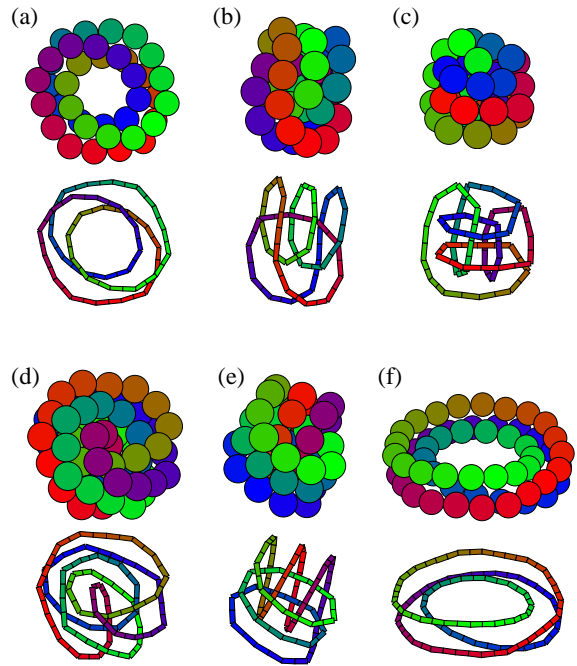}
\caption{(a) St$_{38}$ at $\mu=3.6$: the trefoil knot $3_1$,
(b) St$_{35}$ at $\mu=2.8$: knot $5_1$, (c) St$_{38}$ at $\mu=1.6$: knot $8_{19}$,
(d) St$_{55}$ at $\mu=3.2$: knot $10_{124}$, (e) St$_{45}$ at $\mu=2.6$: knot $10_{139}$,
(f) St$_{54}$ at $\mu=4.6$: a coil with the topology of the trivial knot.
In each case, spherical particles are shown in the upper panel, and the underlying chain of
dipoles in the lower.  The color changes smoothly along the chain.
\label{knots} }
\end{figure}

Analysis of structure in terms of chains allows us to make sense of the intermediate $\mu$
regime, where neither the isotropic LJ energy nor the dipolar contribution
dominates.  One unexpected solution to the frustration between ring formation and
condensation is illustrated in Fig.~\ref{knots}(a).  In this St$_{38}$ cluster, all
the particles belong to a single closed-loop chain, but the chain has the topology of
a non-trivial knot; it cannot be unravelled into a simple ring without
breaking the chain.  At every point in the knot, three portions
of the chain are in close proximity, giving a large number of nearest neighbor interactions
(114 pairs, based on a distance threshold of $1.35\sigma$).  At the same
time, the structure remains open, limiting the bending energy, and keeping unfavorable
combinations of dipole orientations apart.  The St$_{38}$ example has $D_2$ point
group symmetry, i.e. three distinct $C_2$ rotation axes but no mirror planes.
Although the projection may be unfamiliar,
the structure is a trefoil, the simplest non-trivial knot, and is written $3_1$ in
Rolfsen's notation,\cite{Rolfsen76a} meaning the first (in fact, only) knot whose reduced
projection has three crossings.

The topology of a closed-loop chain is invariant to any deformation (ambient isotopy)
that does not break the chain.  In contrast, the occurrence of the trefoil as a global
potential energy minimum clearly relies on the detailed arrangement of the particles.
Nevertheless, it is natural to ask how prevalent the trefoil is, and whether knots
of different topology also
arise.  A reliable way to distinguish prime knots is through their Jones
polynomials,\cite{Jones85a} which are invariant to ambient isotopies and unique for
knots of up to nine crossings, as well as most of ten.  We find the Jones polynomial of
a knot cluster starting from an arbitrary projection of the underlying chain onto two
dimensions.  The so-called bracket polynomial is first derived by considering
all combinations of splits at the crossings in the projection, and is then combined
with a factor depending on the writhe to give the Jones
polynomial.\cite{Lickorish88a,Adams04a}  The corresponding Rolfsen notation\cite{Rolfsen76a}
can then be looked up in tables.\cite{Jones85a,Adams04a}
\par
The trefoil knot, of which Fig.~\ref{knots}(a) is an example,
first appears at $N=21$, though it is necessarily more compact at these
smaller sizes.  This topology dominates the first band
of knots visible in Fig.~\ref{map} in the range $21\le N\le38$.  However,
towards the larger end of this range, a second topology, $5_1$, of greater complexity
appears.  Figure \ref{knots}(b) shows that this class of global
minimum is more compact than the larger trefoils.  It has three neatly stacked turns,
with a thread through the central axis.  The significant bending of the dipole chain is
compensated by the compactness of the overall structure.
\par
Figure \ref{map} shows a second band of knots, starting at lower $\mu$ and rising with
$N$ in the range $38\le N\le55$.  The trefoil does not occur here, but the $5_1$ knot is
a common feature.  We also observe two substantially more intricate knots whose
reduced projections contain ten crossings.  Fortunately, despite having
more than nine crossings, neither has an
ambiguous Jones polynomial,\cite{Jones87a} allowing both to be identified
as described above.  As shown in Fig.~\ref{knots}(e),
one of them, $10_{139}$, resembles the $5_1$ knot in its packing.  The slightly
wider turns now accommodate two threads, creating two stacks each of three turns.
In the illustrated case of St$_{45}$, the stacks are equivalent, being interchanged
by a $C_2$ operation.  The other 10-crossing knot, $10_{124}$, shown in
Fig.~\ref{knots}(d), more closely resembles the twisted wreath of the trefoil in
Fig.~\ref{knots}(a), but with a denser bundle of four turns.
\par
A knot with eight crossings in the reduced projection, $8_{19}$, shown
in Fig.~\ref{knots}(c), makes an appearance at $N=38$ for values of $\mu$ below
those that produce the trefoil at the same $N$.  The smaller
dipole moment means that the isotropic LJ contribution is more influential,
and this cluster is indeed compact, having 156 nearest neighbor pairs, compared with
114 for the trefoil.  These contacts come at the expense of some sharp bends in the
chain of dipoles.
\par
We note that it is possible for a closed chain of dipoles to possess multiple turns
without passing through its own loops to produce a knot.  Such coiled structures are
also observed as global minima, as illustrated by the St$_{54}$ cluster in
Fig.~\ref{knots}(f).  Furthermore, there is the possibility of forming more than one closed
chain in the same structure.  Such combinations are known as links, and we have
already seen a 
topologically trivial example in Fig.~\ref{links}(a).  However, non-trivial links
are also encountered in the structural map of Fig.~\ref{map}.  The smallest and
simplest example is the Hopf link, $2^2_1$, of two interlocked rings, which first
occurs for St$_{12}$ in the range $1.6\le\mu\le2.3$ and consists of two interlocked
hexagons.  A less clear example is illustrated in Fig.~\ref{links}(b) for St$_{48}$,
which is composed of two interlocking coils, each of two turns, and has overall
$C_2$ symmetry.  One three-component link topology, $6^3_3$, consisting of three
mutually interlocked rings, has also been observed, and is illustrated in
Fig.~\ref{links}(c).  The interplay of factors that determine the number of
components in a link is expected to be rather delicate, and indeed St$_{33}$
is a $6^3_3$ link for values of $\mu$ where St$_{32}$ or St$_{34}$ is a knot.
In this case the balance may be tipped by the fact that the 33-particle cluster
can be divided into three rings of equal size.
\par
The occurrence of such topologically exotic structures as global minima of
the simple model Stockmayer potential was unexpected.  On the other hand,
given that knots turned out to be optimal in some cases, it is now
remarkable that only a few topologies are observed for $N\le55$ out
of the 249 possibilities with up to ten crossings.  For larger clusters,
where a single closed-loop chain would be long enough to accommodate even more
crossings, the possibility of yet more complex topologies arises.  However,
the present identification by Jones polynomials would then be
inadequate, since the polynomials are not unique for 10 or more
crossings.\cite{Jones87a}
For sufficiently strong dipole moment, the global minimum of a large
cluster will always be a ring to obtain the maximum number of head-to-tail
contacts.  However, as Fig.~\ref{map} shows, the threshold at which rings
are optimal increases with size.  In contrast, the average dipole strength
at which the LJ structure is superseded shows no overall trend with size.
The range of dipole strengths over which complex structures such as knots
may be found therefore widens with increasing cluster size.
\par
The fact that knots appear over a reasonable spread of sizes and
dipole strengths suggests that there is a good chance of observing
some of them experimentally in suspensions of dipolar colloids.
However, while the advantages of the Stockmayer potential include being well
known, the LJ contribution is only one choice for the isotropic attraction.
For some colloids, a shorter-range potential and stiffer repulsive core
might be more appropriate, and it would be interesting to see whether
the delicate balance of nearest-neighbor and chain-like tendencies
that leads to knots is preserved with respect to such changes.
\par
A number of other questions arise concerning thermodynamic stability
at finite temperature, and the mechanism and dynamics of self-assembly.\cite{Workum05a}
Further work is required to relate these properties to the overall
organization of the underlying potential energy landscape.\cite{Wales04b}
\par
MAM is grateful to Churchill College, Cambridge for financial support.

\end{document}